\documentclass[a4paper,11pt]{article}
\pdfoutput=1 
\newcommand{\comm}[1]{}
\usepackage{xcolor}
\usepackage{jinstpub} 
\usepackage{lineno}
\usepackage{siunitx}
\usepackage{enumitem}
\usepackage{subcaption}
\notoc
\title{\boldmath An FPGA-based Data Aggregator for the New ATLAS ITK Pixel DCS}

\author[a]{A. Qamesh,}
\author[a]{R. Ahmad,}
\author[b]{M. Karagounis,}
\author[a]{P. Kind,}
\author[c]{T. Krawutschke,}
\author[a]{F. Nitz,}
\author[b]{L. Schreiter,}
\author[a]{C. Zeitnitz,}
\affiliation[a]{University of Wuppertal,\\Fakultät 4, \\Gaussstr. 20, 42119 Wuppertal, Germany}
\affiliation[b]{Fachhochschule Dortmund - University of Applied Sciences and Arts ,\\Sonnenstr. 96, 44139 Dortmund, Germany}
\affiliation[c]{Technische Hochschule Köln,\\Gustav-Heinemann-Ufer 54, 50968 Köln}
\emailAdd{ahmed.qamesh@cern.ch}

\abstract{The upcoming ATLAS Phase II upgrade mandates replacing the tracking system with the all-silicon Inner Tracker (ITK), featuring a pixel detector as its core element. The monitoring data of the new system will be aggregated from an on-detector ASIC, Monitoring Of Pixel System (MOPS), and channeled to the Detector Control System (DCS) via a newly developed FPGA-based interface known as MOPS-Hub. 

The mitigation strategies for addressing potential Single-Event Upset (SEU) issues in the new system, along with the proton irradiation results, are presented.}


\collaboration{on behalf of ATLAS ITK collaboration}

\proceeding{Topical Workshop on Electronics for Particle Physics- TWEPP$2024$\\
  30 September to 4 October\\
  Glasgow, UK}
\begin{document}
\maketitle
\flushbottom
\section{Introduction}
\label{sec:introduction}
The planned High Luminosity upgrade of the Large Hadron Collider (HL-LHC) at CERN will increase the collider’s luminosity by a factor of ten compared to the LHC's current luminosity of $10^{34}$ \SI{}{\per\cm\squared\per\second}. As part of the ATLAS Phase II upgrade, the existing tracking system will be replaced with the all-silicon Inner Tracker (ITK), which features a pixel detector as its core element. Monitoring data from the ITK system will be aggregated by an on-detector Application Specific Integrated Circuit (ASIC), known as the Monitoring Of Pixel System (MOPS)~\cite{Walsemann,ahmad2022monitoring}, and relayed to the Detector Control System (DCS) via a newly developed FPGA-based interface known as MOPS-Hub~\cite{Qamesh_2023}. One MOPS chip can monitor temperature and voltage of up to 16 detector modules in a serial power chain~\cite{Walsemann}. 

\begin{figure}[htbp]
\centering
\includegraphics[width=0.7\textwidth]{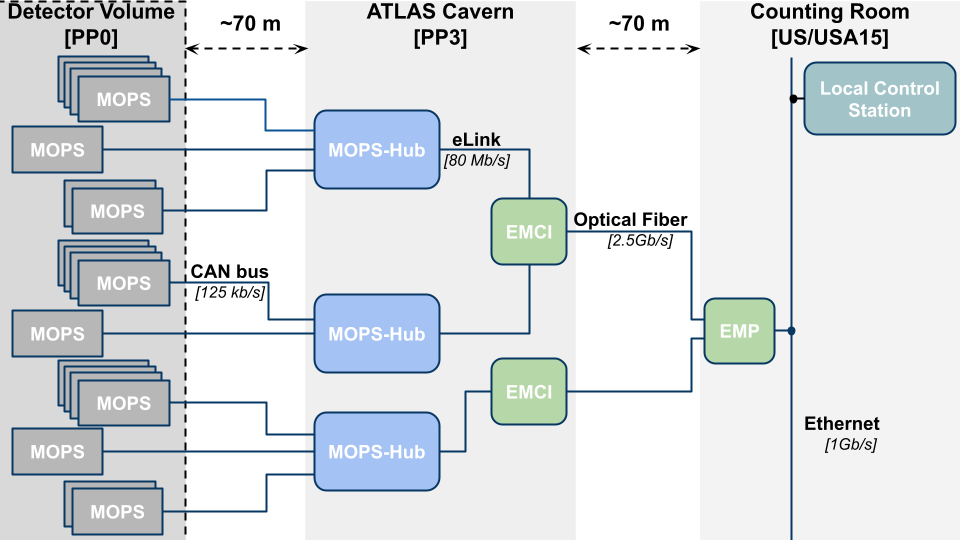}
\caption{\label{fig:construction} The complete MOPS-Hub network.}
\end{figure}

The MOPS-Hub serves as a critical interface, enabling communication with the MOPS chips via CAN bus interfaces and delivering the necessary power through components housed within the MOPS-Hub crate. This crate will operate in racks on the walls of the ATLAS cavern, known as the Patch Panel 3 (PP3). Each MOPS-Hub crate interfaces with 32 CAN buses via CAN Interface Cards (CIC), with one FPGA managing 16 of these buses, referred to as the PP3-FPGA module. Data collected from the CAN buses is forwarded through Electrical Link lines (eLink) to external ATLAS standard DCS components, including Embedded Monitoring and Control Interfaces (EMCIs) and the Embedded Monitoring Processor (EMP) in one chain~\cite{emci_emp}. 
A simplified block diagram of the MOPS-Hub network is depicted in Figure~\ref{fig:construction}.

\section{Expected Radiation Environment and Effects}
\label{sec:radiation_tolerance}
\subsection{Radiation Levels at the ATLAS Cavern}
Despite the location of PP3 outside the ATLAS detector, positioned on the walls of the ATLAS cavern, it still experiences radiation levels.  Table~\ref{tab:rad_tolerance} summarizes the expected radiation level at the walls of the ATLAS cavern  based on the radiation background simulation results in the ATLAS hall using FLUKA (The simulation results are provided by the ATLAS collaboration)~\cite{Dawson:2021xku}. 
\begin{table}[htbp]
\centering
\caption{\label{tab:rad_tolerance} Expected radiation at racks on the walls of the ATLAS cavern~\cite{Dawson:2021xku}. A safety factor (SF) of 3 is applied to the simulated radiation levels.}
\adjustbox{width = 0.8\textwidth}{
\begin{tabular}{l|cc}
\hline
 Parameter      &Expected Dose &Expected Dose with~SF \\
\hline
TID       &  ($\approx$ \SI{30}{\gray})& ($\approx$ \SI{90}{\gray})\\
Neutron fluence ($\Phi_{\text{eq}}^{\text{Si}}$)& \SI{3e11}{N_{eq}\per\cm^2}&\SI{1e12}{N_{eq}\per\cm^2}\\
Hadron fluence >\SI{20}{\mega\eV} ($\Phi_{20}^{\text{Had}}$) &  \SI{2e-7}{/\cm^{2}\per{pp}}& \SI{6e-7}{/\cm^{2}\per{pp}}\\
\hline
\end{tabular}
}
\end{table}

It should be noted that the values in Table~\ref{tab:rad_tolerance} represent an extreme condition, as the PP3 location is situated over 10 meters away from the ATLAS cavern, resulting in lower radiation levels than those indicated.

A critical design challenge is to guarantee that the MOPS-Hub can operate reliably in the harsh radiation environment at PP3. Thus, all commercial
off-the-shelf components  (COTS) on the MOPS-Hub are chosen for their radiation tolerance. Several studies detailed in~\cite{lee2016analysis,lee2014single,cannon2015evaluating,Wirthlin_Kintex,xilinx_ug116} have provided comprehensive reviews of radiation-induced effects in modern FPGAs and the associated mitigation strategies, which informed the implementation of robust mitigation techniques in the Artix-7 FPGA used in the PP3-FPGA module.
\subsection{Radiation effects on the PP3-FPGA module}
Radiation effects in FPGAs can be broadly classified into two categories: Cumulative Effects and Single Event Effects (SEE). 

\subsubsection{Cumulative Effects}
Cumulative effects can be further divided based on the type of incident particles: Total Ionizing Dose (TID) and Displacement Damage (Non-Ionizing Energy Loss: NIEL). Previous TID irradiation campaigns on the Artix-7 FPGA have reached values of \SI{550}{\kilo\radian}\footnote{\SI{100}{\radian} = \SI{1}{\gray}}~\cite{Hu2019}, which is well beyond the expected TID levels shown in Table~\ref{tab:rad_tolerance}. A TID test was conducted on the PP3-FPGA module at the Gamma Irradiation Facility (GIF++)~\cite{pfeiffer2017radiation}, reaching a steady-state TID of \SI{168}{\gray}, with  no abnormal behavior or degradation observed. 

A neutron irradiation test on the PP3-FPGA module was carried out at the TRIGA Reactor at the Jozef Stefan Institute (JSI), Ljubljana, exposing the device to a neutron fluence of $10^{12}$ \si{n/cm^2}. The results indicated no abnormal behavior due to the irradiation. Detailed analysis of these tests is beyond the scope of this work.

\subsubsection{Single event effects}
SEEs are caused by a very high-energy deposition in a small volume of the electronic circuit. The charge generated along the path of an ionizing particle, or a portion of it, may be collected at a microcircuit node, leading to effects such as transient current pulses, change in memory values (bit flips or SEUs), or latchup events, commonly referred to as Single Event Latchup (SEL). In an FPGA, these effects can occur in fabric logic elements, Configuration Memory (CRAM), and internal proprietary control elements, potentially disrupting the FPGA's functionality and causing unexpected system behavior. 
\section{Test Procedure for SEE Characterization}
\label{sec:seu_test_pp3-fpga_procedure}
\begin{figure}[htbp]
\centering
\includegraphics[width=0.9\textwidth]{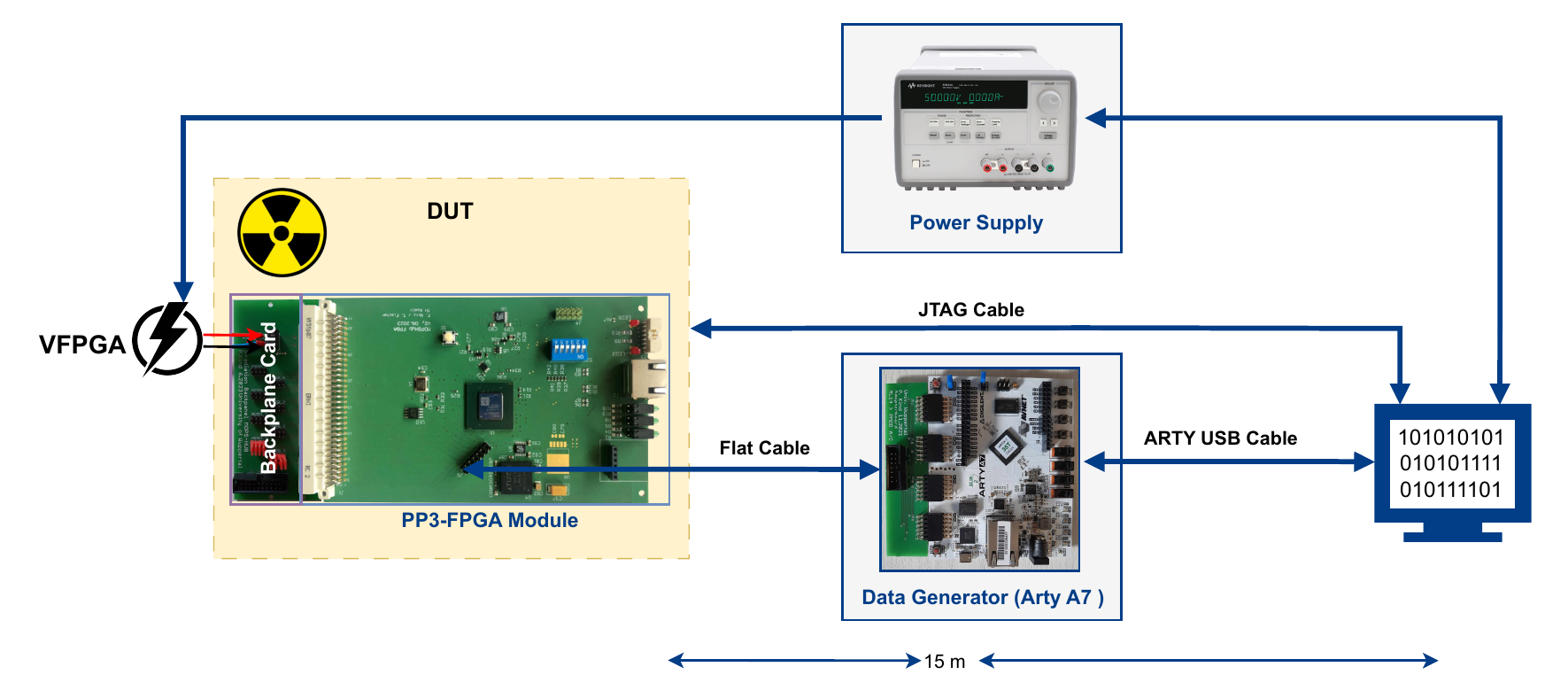}
\caption{Test setup during proton beam testing at HIT facility.}
\label{fig:mopshub_irradiation_test_setup_fpga_seu}
\end{figure}

The SEE characterization of the \textbf{XC7A200T-FPGA} on the PP3-FPGA module, was conducted using the proton facility at the Heidelberg Ion Beam Therapy Center (HIT) in Germany. This facility can accelerate protons up to \SI{221}{\mega\electronvolt} and deliver a beam intensity of up to \SI{3.2e9}{protons/\second}.
Figure~\ref{fig:mopshub_irradiation_test_setup_fpga_seu} depicts the schematics of the test setup during proton beam testing at HIT facility.

In order to estimate the SEU cross-section in the CRAM, a test logic consisting of a \SI{3000}{\bit}-long shift register was implemented in the FPGA firmware. The \textbf{ARTY board} depicted in the setup acts as a control unit, writing to and reading data from the \SI{3000}{\bit} shift register to check for any mismatch. The Device Under Test (DUT) was aligned at the center of the beam to ensure that only the $23 \times 23$ \SI{}{\mm} XC7A200T-FPGA was irradiated, without affecting other electronic components on the PP3-FPGA module.

The PP3-FPGA module operated at a constant voltage  ($\text{V}_\text{FPGA}$ = \SI{5}{\volt}), ensuring a stable power supply during radiation exposure. Three copies of the firmware were stored in the external radiation hard flash memory.  To avoid the destruction of the DUT due to SELs, the DUT power supply current was continuously monitored from the a Host Computer using a python script. Firmware reloading via JTAG was only considered during the radiation campaign in cases where all copies in the flash memory were corrupted. Two distinct operational scenarios were considered for this test.

First scenario: The Multi-boot Auto Reconfiguration (mBAR) mechanism was evaluated with an external supervisory watchdog enabled\footnote{TPS3306 Watchdog/Supervisory IC is used~\cite{watchdog_tps3306}} to hold the FPGA in reset as long as the supply voltage is insufficient~\cite{watchdog_tps3306}. During operation, if the FPGA fails to send a heartbeat within \SI{0.8}{\second}, the watchdog resets the FPGA, and a new configuration is loaded from the external radiation-hard memory. Under this condition, the DUT was powered while the Xilinx Soft Error Mitigation (SEM) IP was employed to monitor the configuration RAM (CRAM) and automatically correct any detected errors.  

Second scenario: The external watchdog was intentionally disabled to prevent unnecessary resets during the read/write process of the implemented shift register. This allowed for uninterrupted measurements during SEU cross-section  calculations for the XC7A200T-FPGA under radiation exposure.

\section{Summary of the Experimental Results}
\label{sec:seu_test_pp3_fpga_campaign}

\subsection{CRAM Behaviour}
\label{sec:seu_test_pp3_fpga_cram_behaviour}
The proton irradiation of the PP3-FPGA module with the watchdog enabled was performed with varying fluences for 6 runs, reaching a maximum fluence of \SI{7.5e12}{protons\per\cm\squared} during the campaign.
During irradiation, the average reset rate due to mBAR was observed to be up to \SI{2}{resets/\second}. This elevated reset rate is attributed to a high upset rate in the CRAM. While some upsets are correctable by the SEM IP, there remains a risk of Multi Bit Upset (MBU) affecting more than two bits per frame simultaneously. Such events can overwhelm the protection mechanisms of the SEM IP and cannot be corrected~\cite{Xilinx2017SEM}. Accumulation of upsets, particularly in critical areas in critical circuits such as clock generators, can break the system’s operation. Unrecoverable errors cannot be repaired without re-downloading the FPGA firmware using the mBAR mechanism.
\subsection{ SEU Estimation}
\label{sec:pp3_fpga_rad_metrics}
SEU estimation was performed under varying fluences, with the watchdog disabled, reaching a maximum of \SI{2e12}{protons/cm^2}, approximately \SI{e4}{} times the expected fluence at the PP3 location mentioned in Table~\ref{tab:rad_tolerance}.

With the help of the ARTY Board depicted in Figure~\ref{fig:mopshub_irradiation_test_setup_fpga_seu}, a specific pattern was shifted through the shift register in the PP3-FPGA module at \SI{1}{\kilo\hertz} clock rate, with a hold time of \SI{1}{\second} and then read back. When a mismatch was found, the XC7A200T-FPGA reported an SEU failure, incremented the failure counter in the ARTY Board, and sent the information to the Host computer for diagnosis. The Run details summary for the campaign results are listed in Table~\ref{tab:runs_seus}, where the SEU cross-sections align with other measurements conducted by Xilinx on the same \textbf{Artix-7} family~\cite{xilinx_ug116}.
\begin{table}[htbp]
\centering
\caption{Run details summary for the PP3-FPGA with the watchdog disabled, during the proton campaign.}
\label{tab:runs_seus}
\adjustbox{width = 0.9\textwidth}{
\begin{tabular}{cccccccc}
\hline
Run & Energy&FWHM &Intensity&Duration&  Fluence& SEU &$\sigma$\\

 &[MeV] & [mm]&[protons/s]  &[min]&[protons/cm$^2$] &  [N]& [cm$^2$/bit]\\

\hline
7  & 145.46 &11.5& \num{8e8} &37 &\num{8e11} & 7 &$2.92\times$\num{e-15}\\

8  & 155.82 &10.8& \num{8e8} &39&\num{8e11} & 9  & $3.75\times$\num{e-15}\\

9  & 165.89 &10.2& \num{8e8} &17 &\num{4e11} &4  &

$3.33\times$\num{e-15}\\ 

\hline
\end{tabular}
}
\end{table}

As detailed in Table~\ref{tab:runs_seus}, the cross-section is energy independent within the uncertainty. These results agree with the simulation studies for FPGA devices for the upset rates at different  energies discussed in~\cite{Faccio_computational}. The estimated rate of SEU in the CRAM of the~\textbf{XC7A200T-FPGA}\footnote{Total RAM Bits in XC7A200T-FPGA = 13455360 bits} at PP3-location, based on Equation~\ref{eq:sigma_seu} can reach up to 25349 SEUs in 10 years of the LHC operation, assuming the maximum~SEUs cross-section estimated in Table~\ref{tab:runs_seus}.  
\begin{equation}
\label{eq:sigma_seu}
{\sigma_{SEU}}= \frac{\sigma_{SEU}}{N_{bits}} =\frac{N_{SEU}}{\phi\times N_{bits}}\hspace{1cm} [\SI{}{\cm^2}]
\end{equation}

Where $N_{SEU}$ is the number of SEUs that occur, ${\phi}$ is the total incident particle Fluence estimated with that of hadrons above \SI{20}{\mega\eV} as detailed in Table~\ref{tab:rad_tolerance}. To mitigate the impact of SEUs, the firmware design of the PP3-FPGA incorporates the TMR technique, which minimizes radiation-induced errors and enhances data reliability described in~\cite{Qamesh_2023}.
\subsection{Current and Voltage Behavior}
\label{sec:seu_test_pp3_fpga_current}
A preset limit on the supply current (I$_\text{supply}$) was established during the irradiation campaign at \SI{510}{\milli\ampere}, which is 10\% above the calculated static current mentioned in~\cite{artix7_datasheet} to safeguard the FPGA against potential over-current scenarios caused by SELs.

\begin{figure}[htbp]
  \centering

    \begin{subfigure}[b]{0.49\textwidth}
    \centering
    \includegraphics[width=\textwidth, height=0.2\textheight]{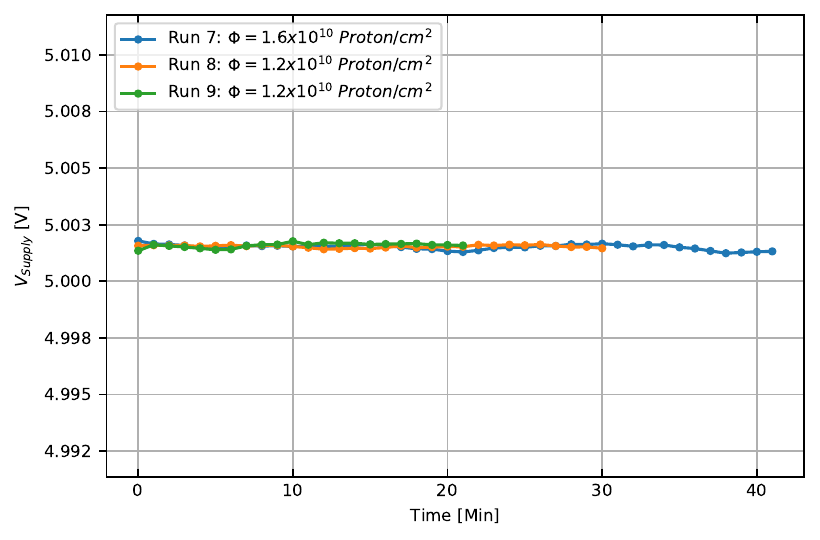}
    \caption{Supply voltage during proton irradiation.}
    \label{fig:pp3_fpga_seu_voltage}
  \end{subfigure}
  \hfill
  \begin{subfigure}[b]{0.49\textwidth}
    \centering
    \includegraphics[width=\textwidth, height=0.2\textheight]{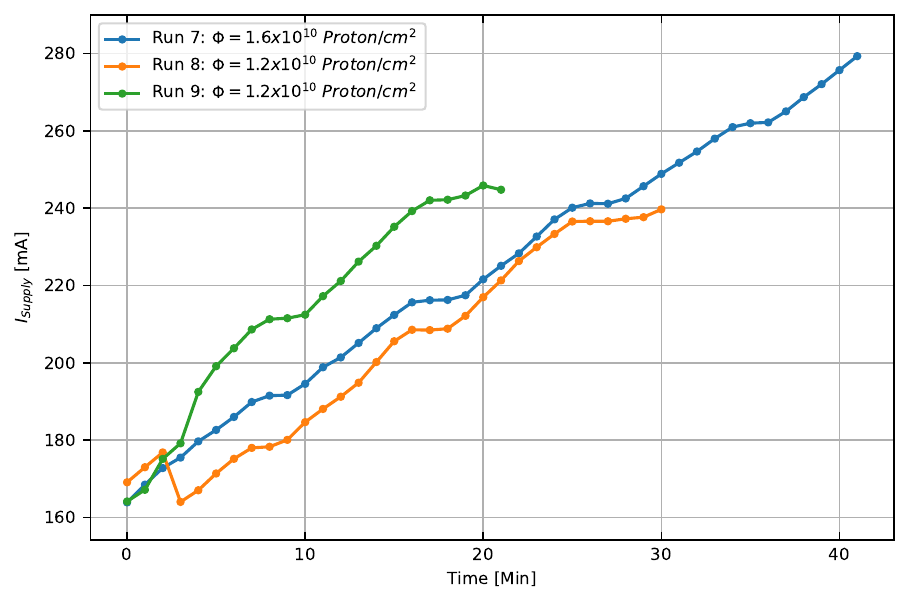}
    \caption{Supply current during proton irradiation.}
    \label{fig:pp3_fpga_seu_current}
  \end{subfigure}
  \caption{Supply power to the PP3-FPGA module, with the watchdog disabled,  during the proton campaign.}
  \label{fig:pp3_fpga_seus_electrical}
\end{figure}

Throughout the irradiation process, no sudden increases in the power supply current were observed. This stability suggests that no latchups occurred, which typically cause abrupt current spikes. However, it was noted that as SEU accumulated, there was a gradual increase in the current consumption of the FPGA, as depicted in Figure~\ref{fig:pp3_fpga_seu_current}. Immediately after reconfiguration was enabled, the current dropped back to its initial level. This phenomenon is consistent with observations reported in other SEU studies~\cite{ZBTSRAM} and shouldn't cause an issue in the final system since the accumulation of such errors is managed using the mBAR approach discussed in Section~\ref{sec:seu_test_pp3_fpga_cram_behaviour}. 

\section{Conclusion}
\label{sec:summary}
The PP3-FPGA module is a critical component in the data aggregation of the MOPS-Hub crate.  It utilizes a commercial Xilinx Artix-7 XC7A200T-FPGA which is vulnerable to SEEs in the PP3 environment. The FPGA performance was demonstrated at the proton facility in HIT.  The assessment of the SEU rate detailed in the previous sections for the XC7A200T-FPGA reveals a calculated rate of approximately 7 upsets per day in the PP3 environment. This effect is mitigated in the final system through a combination of both circuit level (TMR and SEM IP) and system level (mBAR) strategies.

\end{document}